\documentclass[useAMS,usenatbib]{mn2e}
\usepackage{graphics}
\usepackage{graphicx}
\usepackage{psfig}
\input{epsf}
\def\l{\Lambda}

\def\etal{et al.}

\def \lleq {\lower0.9ex\hbox{ $\buildrel < \over \sim$} ~}
\def \ggeq {\lower0.9ex\hbox{ $\buildrel > \over \sim$} ~}

\def\omt{\Omega_{0 \rm m}}

\def\spose#1{\hbox to 0pt{#1\hss}}
\def\simle{\mathrel{\spose{\lower 3pt\hbox{$\mathchar"218$}}
     \raise 2.0pt\hbox{$\mathchar"13C$}}}
\def\simge{\mathrel{\spose{\lower 3pt\hbox{$\mathchar"218$}}
     \raise 2.0pt\hbox{$\mathchar"13E$}}}

\def\apj{Astroph.~J.~}
\def\mn{Mon.~Not.~Roy.~Ast.~Soc.~}

\def\aj{Astron.~J.~}
\def\prl{Phys.~Rev.~Lett.~}
\def\prd{Phys.~Rev.~D~}

\def\plb{Phys.~Lett.~B~}
\def\jetpl{JETP ~Lett.~}

\def\beq{\begin{equation}}
\def\eeq{\end{equation}}
\def\ber{\begin{eqnarray}}
\def\eer{\end{eqnarray}}

\begin{document}

\title{Smoothing Supernova Data to Reconstruct the Expansion History
of the Universe and its Age}

\author[Arman Shafieloo, Ujjaini Alam, Varun Sahni, Alexei A. Starobinsky]{Arman Shafieloo$^{1,3}$, Ujjaini Alam$^{1,4}$, Varun Sahni$^{1,5}$ and Alexei A. Starobinsky$^{2,6}$ \\
  $^{1}$ Inter University Centre for Astronomy $\&$ Astrophysics, Pune, India\\
  $^{2}$ Landau Institute for Theoretical Physics, 119334 Moscow, Russia\\
  $^{3}$arman@iucaa.ernet.in \\
  $^{4}$ujjaini@iucaa.ernet.in \\
  $^{5}$varun@iucaa.ernet.in \\
  $^{6}$alstar@landau.ac.ru \\
  } \maketitle

\begin{abstract}
We propose a non-parametric method of smoothing supernova data over
redshift using a Gaussian kernel in order to reconstruct
important cosmological quantities including $H(z)$ and $w(z)$ in a
model independent manner.  This method is shown to be successful in
discriminating between different models of dark energy when the
quality of data is commensurate with that expected from the future
SuperNova Acceleration Probe (SNAP). We find that the Hubble parameter
is especially well-determined and useful for this purpose. The look
back time of the universe may also be determined to a very high degree
of accuracy ($\lleq 0.2 \%$) in this method. By refining the method,
it is also possible to obtain reasonable bounds on the equation of
state of dark energy.  We explore a new diagnostic of dark energy--
the `$w${\tt -probe}'-- which can be calculated from the first derivative of
the data. We find that this diagnostic is reconstructed extremely
accurately for different reconstruction methods even if $\omt$ is
marginalized over. The $w${\tt -probe} can be used to successfully
distinguish between $\Lambda$CDM and other models of dark energy to a
high degree of accuracy.
\end{abstract}


\section{Introduction}

The nature of dark energy has been the subject of much debate over the
past decade (for reviews see
\cite{ss00,carroll01,pr02,paddy03,sahni04}). The supernova (SNe) 
type Ia data, which gave the first indications of the accelerated 
expansion of the universe, are expected to throw further light on this
intriguing question as their quality steadily improves. While the
number of SNe available to us has increased two-fold over the past
couple of years (at present there are about $150$ SNe between
redshifts of $0$ and $1.75$, with $10$ SNe above a redshift of unity)
\citep{riess,perl,knop,tonry03,hst}, the SNe data are still not of a quality
to firmly distinguish different models of dark energy. In this
connection, an important role in our quest for a deeper understanding
of the nature of dark energy has been played by the `reconstruction
program'. Commencing from the first theoretical exposition of the
reconstruction idea -- \cite{St98,HT99,NC99}, and \cite{SSSS} which
applied it to an early supernova data set-- there have been many
attempts to reconstruct the properties of dark energy directly from
observational data without assuming any particular
microscopic/phenomenological model for the former. When using SNe data
for this purpose, the main obstacle is the necessity to: (i)
differentiate the data once to pass from the luminosity distance $d_L$
to the Hubble parameter $H(t)\equiv \dot a(t)/a(t)$ and to the
effective energy density of dark energy $\epsilon_{DE}$, (ii)
differentiate the data a second time in order to obtain the
deceleration parameter $q\equiv -\ddot a a/\dot a^2$, the dark energy
effective pressure $p_{DE}$, and the equation of state parameter
$w(t)\equiv p_{DE}/\epsilon_{DE}$. Here, $a(t)$ is the scale factor of
a Friedmann-Robertson-Walker (FRW) isotropic cosmological model which
we further assume to be spatially flat, as predicted by the simplest
variants of the inflationary scenario of the early Universe and
confirmed by observational CMB data.

To get around this obstacle, some kind of smoothing of $d_L$ data with
respect to its argument -- the redshift $z(t)$ -- is needed.  One
possible way is to parameterize the quantity which is of interest
($H(z)$, $w(z)$, etc.) by some functional form containing a few free
parameters and then determine the value of these parameters which
produce the best fit to the data.  This implies an implicit smoothing
of $d_L$ with a characteristic smoothing scale defined by the number
of parameters, and with a weight depending on the form of
parameterization.  Different parameterizations have been used for:
$d_L$ \citep{HT99,SSSS,chiba}, $H(z)$ \citep{rs1,alam03b, alam04a},
$w(z)$ \citep{polar,albrecht,efstathiou,maor02,copeland,linder,wangm,
saini,leandros,gong,Lazkoz} and $V(z)$ \citep{simon,ohta}. In
\cite{HT99}, a polynomial expansion of the luminosity distance was
used to reconstruct the equation of state. However, \cite{albrecht}
showed this ansatz to be inadequate since it needed an arbitrarily 
large number of parameters to fit even the simplest $\Lambda$CDM 
equation of state. They proposed instead a polynomial ansatz for
the equation of state which worked somewhat better. In \cite{SSSS} a
rational Pad{\`e}-type ansatz for $d_L$ was proposed, which gave good
results.  In recent times there have been many more attempts at
parameterizing dark energy.  In \cite{polar} and \cite{linder} an
ansatz of the form $w=w_0+w_a(1-a)$ was suggested for the equation of
state.  \cite{copeland} suggested a four-parameter ansatz for the
equation of state. \cite{rs1} proposed a slightly different approach
in which the dark energy density was expanded in a polynomial ansatz,
the properties of which were examined in \citep{alam03b,
  alam04a,alam04b}.  See \cite{alam03a, gong1,bass} for a summary of
different approaches to the reconstruction program and for a more
extensive list of references. In spite of some ambiguity in the form
of these different parameterizations, it is reassuring that they
produce consistent results for the best fit curve over the
range $0.1 \lleq z \lleq 1$ where we have sufficient amount of data
(see, e.g., Fig. 10 in \cite{gong1}). However it is necessary to point
out that the current SNe data are not of a quality that could allow us
to unambiguously differentiate $\Lambda$CDM from evolving dark energy.
That is why our focus in this paper will be on better quality data
(from the SNAP experiment) which should be able to successfully
address this important issue.

A different, non-parametric smoothing procedure involves directly
smoothing either $d_L$, or any other quantity defined within redshifts
bins, with some characteristic smoothing scale. Different forms of
this approach have been elaborated in
\cite{wang01,hut03,saini03,daly03,daly04,wangteg05,inverse}. One of
the advantages of this approach is that the dependence of the results
on the size of the smoothing scale becomes explicit. We emphasize
again that the present consensus seems to be that, while the
cosmological constant remains a good fit to the data, more exotic
models of dark energy are by no means ruled out (though their
diversity has been significantly narrowed already). Thus, until the
quality of data improves dramatically, the final judgment on the
nature of dark energy cannot yet be pronounced.

In this paper, we develop a new reconstruction method which formally
belongs to the second category, and which is complementary to the
approach of fitting a parametric ansatz to the dark energy density or
the equation of state.  Most of the papers using the non-parametric
approach cited above exploited a kind of top-hat smoothing in redshift
space.  Instead, we follow a procedure which is well known and
frequently used in the analysis of large-scale structure
\citep{coles,martinez02}; namely, we attempt to smooth noisy data
directly using a Gaussian smoothing function. Then, from the smoothed
data, we calculate different cosmological functions and, thus, extract
information about dark energy. 
This method allows us to avoid additional noise due to sharp borders
between bins. Furthermore, since our method
does not assume any
definite parametric representation of dark energy, it does
not bias results towards any particular model. 
We therefore expect this method to give us model-independent
estimates of cosmological functions, in particular, the Hubble
parameter $H(z)\equiv \dot a(t)/a(t)$. On the basis of data
expected from the SNAP satellite mission, we show that the Gaussian
smoothing ansatz proposed in this paper can successfully distinguish
between rival cosmological models and help shed light on the nature of
dark energy.

\begin{table*}
\centering
\caption{Expected number of supernovae per redshift bin from
the SNAP experiment}
\label{tab:SNAP}
\footnotesize
\begin{center}
\begin{tabular}{ccccccccc}
\hline
\hline
$\Delta z$&0.1--0.2&0.2--0.3&0.3--0.4&0.4--0.5&0.5--0.6&0.6--0.7&0.7--0.8&0.8--0.9\\
\hline
$N$&35&64&95&124&150&171&183&179\\
\hline
&&&&&&&&\\
\hline
$\Delta z$&0.9--1.0&1.0--1.1&1.1--1.2&1.2--1.3&1.3--1.4&1.4--1.5&1.5--1.6&1.6--1.7\\
\hline
$N$&170&155&142&130&119&107&94&80\\
\hline
\end{tabular}
\end{center}
\end{table*}

\section{Methodology}\label{sec:method}

It is useful to recall that, in the context of structure formation, it
is often advantageous to obtain a smoothed density field
$\delta^S({\bf x})$ from a fluctuating `raw' density field,
$\delta({\bf x'})$, using a low pass filter $F$ having a
characteristic scale $R_f$ \citep{coles}
\beq
\delta^S({\bf x}, R_f) = \int \delta({\bf x'}) F(|{\bf x}-{\bf x'}|;R_f)~d{\bf x'}~.
\eeq
Commonly used filters include: (i) the `top-hat' filter, which has a
sharp cutoff $F_{\rm TH} \propto \Theta(1-|{\bf x}-{\bf x'}|/R_{\rm
  TH})$, where $\Theta$ is the Heaviside step function ($\Theta(z) =
0$ for $z \leq 0$, $\Theta(z) = 1$ for $z > 0$) and (ii) the Gaussian
filter $F_{\rm G} \propto \exp(-|{\bf x}-{\bf x'}|^2/2R_{\rm G}^2)$.
For our purpose, we shall find it useful to apply a variant of the
Gaussian filter to reconstruct the properties of dark energy from
supernova data. In other words, we apply Gaussian smoothing to
supernova data (which is of the form $\lbrace \ln d_L(z_i),z_i
\rbrace$) in order to extract information about important cosmological
parameters such as $H(z)$ and $w(z)$. The smoothing algorithm
calculates the luminosity distance at any arbitrary redshift $z$ to be
\ber
\label{eq:bg}
\ln d_L(z,\Delta)^{\rm s}=\ln
\ d_L(z)^g+N(z) \sum_i \left [ \ln d_L(z_i)- \ln
\ d_L(z_i)^g \right] &&\nonumber\\
{\large \times} \ {\rm exp} \left [- \frac{\ln^2 \left
( \frac{1+z_i}{1+z} \right ) }{2 \Delta^2} \right ], &&\\
N(z)^{-1}=\sum_i {\rm exp} \left
[- \frac{\ln^2 \left ( \frac{1+z_i}{1+z} \right ) }{2 \Delta^2}
\right ]~. \hspace{2.8cm}&&\nonumber
\eer
Here, $\ln d_L(z,\Delta)^{\rm s}$ is the smoothed luminosity distance
at any redshift $z$ which depends on luminosity distances of each SNe
event with the redshift $z_i$, and $N(z)$ is a normalization
parameter. Note that the form of the kernel bears resemblance to the
lognormal distribution (such distributions find application in the
study of cosmological density perturbations, \cite{colesrev}).  The
quantity $\ln d_L(z)^g$ represents a guessed background model which we
subtract from the data before smoothing it.  This approach allows us
to smooth noise only, and not the luminosity distance. After noise
smoothing, we add back the guess model to recover the luminosity
distance.  This procedure is helpful in reducing noise in the results.
Since we do not know which background model to subtract, we may take a
reasonable guess that the data should be close to $\Lambda$CDM and use
$d_L(z)^g=d_L(z)^{\Lambda {\rm CDM}}$ as a first approximation and then use a
boot-strapping method to find successively better guess models. We
shall discuss this issue in greater detail in the
section~\ref{sec:results}. Having obtained the smoothed luminosity
distance, we differentiate once to obtain the Hubble parameter $H(z)$
and twice to obtain the equation of state of dark energy $w(z)$, using
the formula
\beq
\label{eq:hubb}
H(z)= \left[ \frac{d}{dz} \left( \frac{d_L(z)}{1+z} \right) \right]^{-1}\,\,,
\eeq
\beq
\label{eq:w}
w(z)=\frac{[2 (1+z) /3] \ H^{\prime}/H - 1}{1 \ - \ (H_0/H)^2
\omt \ (1+z)^3} \,\,.
\eeq
The results will clearly depend upon the value of the scale $\Delta$
in (\ref{eq:bg}). A large value of $\Delta$ produces a smooth result,
but the accuracy of reconstruction worsens, while a small $\Delta$
gives a more accurate, but noisy result. Note that, for $|z-z_i|\ll
1$, the exponent in Eq.~(\ref{eq:bg}) reduces to the form
$-(z-z_i)^2/2\Delta^2(1+z)^2$. Thus, the effective Gaussian smoothing
scale for this algorithm is $\Delta (1+z)$. We expect to obtain an
optimum value of $\Delta$ for which both smoothness and accuracy are
reasonable.

The Hubble parameter can also be used to obtained the weighted average
of $w$
\beq
1+\bar{w}=\frac {1}{\delta \ln(1+z)}\int \left(1+w(z)\right)\frac{dz}{1+z}=\frac{1}{3} \frac{\delta \ \ln \tilde{\rho}_{DE}}{\delta \ \ln (1+z)}\,\,.
\eeq
$\tilde {\rho}_{DE}$ is the dark energy density $\tilde {\rho}_{DE}=\rho_{DE}/\rho_{0c}$ (where $\rho_{0c}=3H^2_0/8\pi G$). We shall show in the section~\ref{sec:avgw} that $\bar{w}$, which we
call the $w$-probe, acts as an excellent diagnostic of dark energy,
and can differentiate between different models of dark energy with
greater accuracy than the equation of state.

\begin{figure*}
\begin{center}
\centerline{\mbox{{\large $\Delta=0.24$}}}
\centerline{\mbox{{\normalsize Fiducial Model: $w=-1/(1+z)$}}}
\psfig{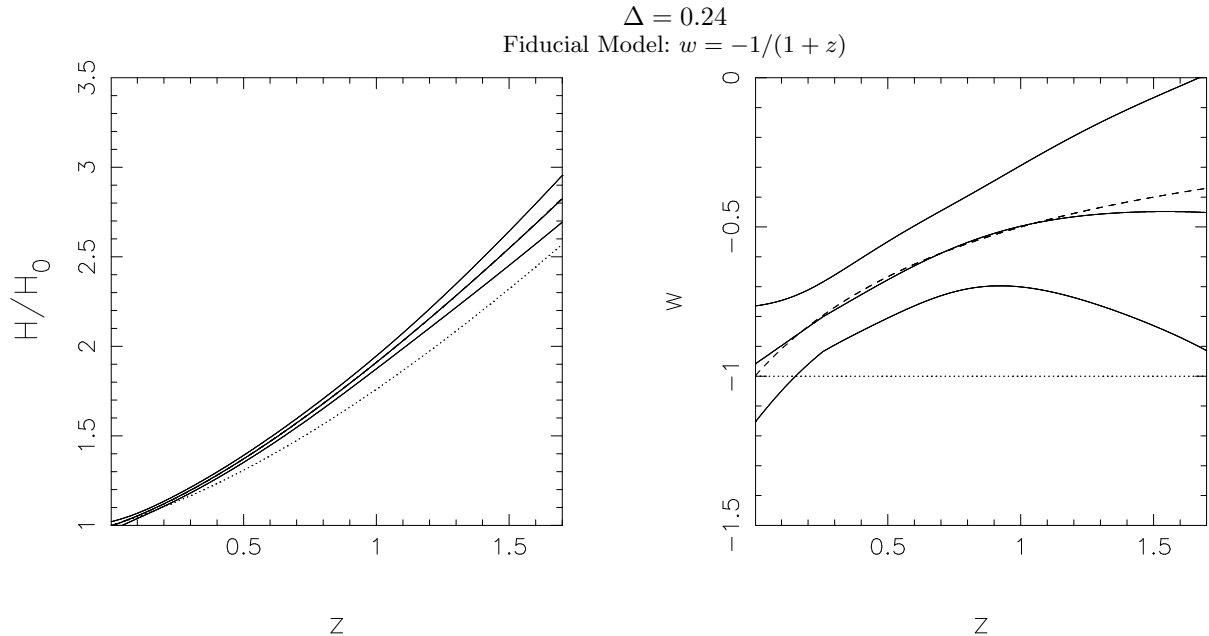}
\vspace{-0.6cm}
\end{center}
\caption{\small The smoothing scheme of equation~(\ref{eq:bg}) is used
to determine $H(z)$ and $w(z)$ from $1000$ realizations of the SNAP
dataset.  The smoothing scale is $\Delta=0.24$.  The dashed line in
each panel represents the fiducial $w=-1/(1+z)$ `metamorphosis' model
while the solid lines show the mean Hubble parameter (left), the mean
equation of state (right), and $1\sigma$ limits around these
quantities.  The dotted line in both panels is $\Lambda$CDM. Note that the
mean Hubble parameter is reconstructed so accurately that the fiducial
model (dashed line) is not visible in the left panel.}
\label{fig:w=var,D=0.24,SNAP}
\end{figure*}

To check our method, we use data simulated according to the SuperNova
Acceleration Probe (SNAP) experiment. This space-based mission is
expected to observe close to $6000$ supernovae, of which
about $2000$ supernovae can be used for cosmological purposes
\citep{snap04}. We propose to use a distribution of $1998$ supernovae
between redshifts of $0.1$ and $1.7$ obtained from \cite{snap04}. This
distribution of $1998$ supernovae is shown in Table ~\ref{tab:SNAP}.
Although SNAP will not be measuring supernovae at redshifts below
$z=0.1$, it is not unreasonable to assume that, by the time SNAP comes
up, we can expect high quality data at low redshifts from other
supernova surveys such as the Nearby SN Factory
\footnote{http://snfactory.lbl.gov}. Hence, in the low redshift region
$z < 0.1$, we add $25$ more supernovae of equivalent errors to the
SNAP distribution, so that our data sample now consists of $2023$
supernovae .  Using this distribution of data, we check whether the
method is successful in reconstructing different cosmological
parameters, and also if it can help discriminate different models of
dark energy.

We simulate $1000$ realizations of data using the SNAP distribution
with the error in the luminosity distance given by $\sigma_{\ln
  d_L}=0.07$ -- the expected error for SNAP. We also consider the
possible effect of weak-lensing on high redshift supernovae by adding
an uncertainty of $\sigma_{lens}(z) \approx 0.46 (0.00311+0.08687
z-0.00950 z^2)$ (as in \cite{wangteg05}).  Initially, we use a simple
model of dark energy when simulating data -- an evolving model of dark
energy with $w=-a/a_{0}=-1/(1+z)$ and $\omt=0.3$. It will clearly be of
interest to see whether this model can be reconstructed accurately and
discriminated from $\Lambda$CDM using this method. From the SNAP
distribution, we obtain smoothed data at $2000$ points taken uniformly
between the minimum and maximum of the distributions used.  Once we
are assured of the efficacy of our method, we shall also attempt to
reconstruct other models of dark energy. Among these, one is the
standard cosmological constant ($\Lambda$CDM) model with $w=-1$. The
other is a model with a constant equation of state, $w=-0.5$. Such
models with constant equation of state are known as quiessence models
of dark energy \citep{alam03a} and we shall refer to this model as the
``quiessence model'' throughout the paper.  These three models are
complementary to each other. For the $\Lambda$CDM model, the equation
of state is constant at $w=-1$, $w$ remains constant at $-0.5$ for the
quiessence model and for the evolving model, $w(z)$ varies rapidly,
increasing in value from $w_0 = -1$ at the present epoch to $w \simeq
0$ at high redshifts.

\section{Results}\label{sec:results}

In this section we show the results obtained when our smoothing scheme
is applied to data expected from the SNAP experiment.  The first issue
we need to consider is that of the guess model. As mentioned earlier,
the guess model in equation~(\ref{eq:bg}) is arbitrary. Using a guess
model will naturally cause the results to be somewhat biased towards
the guess model at low and high redshifts where there is paucity of
data. Therefore we use an iterative method to estimate the guess model
from an initial guess.

\noindent{\bf Iterative process to obtain Guess model}

To estimate the guess model for our smoothing scheme, we use the
following iterative method. We start with a simple cosmological model,
such as $\Lambda$CDM, as our initial guess model-- $\ln d_L^{g0}=\ln
d_L^{\Lambda {\rm CDM}}$. The result obtained from this analysis, $\ln
d_L^1$, is expected to be closer to the real model than the initial
guess. We now use this result as our next guess model-- $\ln
d_L^{g1}=\ln d_L^1$ and obtain the next result $\ln d_L^2$. With each
iteration, we expect the guess model to become more accurate, thus
giving a result that is less and less biased towards the initial guess
model used. A few points about the iterative method should be noted
here.

\begin{itemize}

\item Using different models for the initial guess does not affect the
final result provided the process is iterated several times. For
example, if we use a $w=-1/(1+z)$ `metamorphosis' model to simulate
the data and use either $\Lambda$CDM or the $w=-0.5$ quiessence
model as our initial guess, the results for the two cases converge by
$\ggeq 5$ iterations.

\item Using a very small value of $\Delta$ will result in a
accurate but noisy guess model, therefore after a few iterations, the
result will become too noisy to be of any use. Therefore, we should
use a large $\Delta$ for this process in order to obtain smoother 
results.

\item The bias of the final result will decrease with each iteration,
since with each iteration we get closer to the true model. The bias
decreases non-linearly with the number of iterations $M$. Generally,
after about $10$ iterations, for moderate values of $\Delta$, the bias
is acceptably small. Beyond this, the bias still decreases with the
number of iterations but the decrease is negligible while the process
takes more time and results in larger errors on the parameters.

\item It is important to choose a value of $\Delta$ which gives a small
value of bias and also reasonably small errors on the derived
cosmological parameters.  To estimate the value of $\Delta$ in
(\ref{eq:bg}), we consider the following relation between the
reconstructed results, quality and quantity of the data and the
smoothing parameters. One can show that the
relative error bars on $H(z)$ scale as \citep{tegmark}
\beq
\label{eq:tegmark}
\frac{\delta H}{H}\propto \frac{\sigma}{N^{1/2} \Delta^{3/2}}\,\,,
\eeq 
where N is the total number of supernovae (for approximately uniform
distribution of supernovae over the redshift range) and $\sigma$ is
the noise of the data. From the above equation we see that a larger
number of supernovae or larger width of smoothing, $\Delta$, will
decrease the error bars on reconstructed $H$, but as we shall show in
appendix~\ref{app:errors_bias}, the bias of the method is approximately
related to $\Delta^2$. This implies that, by increasing $\Delta$ we
will also increase the bias of the results.  We attempt to estimate
$\Delta$ such that the error bars on $H$ be of the same order as
$\sigma$, which is a reasonable expectation. 

If we consider a single iteration of our method, then for $N \simeq
2000$ we get $\Delta_{0} \simeq N^{-1/3} \simeq 0.08$.  However, with
each iteration, the errors on the parameters will increase. Therefore
using this value of $\Delta$ when we use an iterative process to find
the guess model will result in such large errors on the cosmological
parameters as to render the reconstruction exercise meaningless.  It
shall be shown in Appendix~\ref{app:errors_bias} that at the M-th
iteration, the error on $\ln d_L$ will be approximately $\delta_M (\ln
d_L) \simeq \sqrt{M} \delta_0 (\ln d_L)$. The error on $\ln d_L$
scales as $1/\Delta$. We would like the errors after $M$ iterations to
be commensurate with the optimum errors obtained for a single
iteration, $\Delta_{0}$, so we require $\Delta_{\rm optimal} \simeq
\sqrt{M} \Delta_{0}$. Therefore, if we wish to stop the boot-strapping
after $10$ iterations, then $\Delta_{\rm optimal} \simeq 3 \Delta_{0}
\simeq 0.24$.  This is the optimal value of $\Delta$ we shall use for
best results for our smoothing procedure.

\end{itemize}

Considering all these factors, we use a smoothing scale $\Delta=0.24$
for the smoothing procedure of Eq~(\ref{eq:bg}) with a iterative 
method for finding the guess model (with $\Lambda$CDM as the initial
guess). The boot-strapping is stopped after $10$ iterations.  We
will see that the results reconstructed using these parameters do not
contain noticeable bias and the errors on the parameters are also
satisfactory.

Figure~\ref{fig:w=var,D=0.24,SNAP} shows the reconstructed $H(z)$ and
$w(z)$ with $1\sigma$ errors for the $w=-1/(1+z)$ evolving model of
dark energy.  From this figure we can see that the Hubble parameter is
reconstructed quite accurately and can successfully be used to
differentiate the model from $\Lambda$CDM.  The equation of state, however,
is somewhat noisier.
There is also a slight bias in the equation of state
at low and high redshifts. Since the $w=-1/(1+z)$ model has an
equation of state which is very close to $w=-1$ at low redshifts, we
see that $w(z)$ cannot discriminate $\Lambda$CDM from the fiducial model at
$z \lleq 0.2$ at the $1 \sigma$ confidence level.

\begin{figure}
\begin{center}
\centerline{\mbox{{\large $\Delta=0.24$}}}
\centerline{\mbox{{\normalsize Fiducial Model: $w=-1/(1+z)$}}}
\vspace{-0.1cm}
\psfig{figure=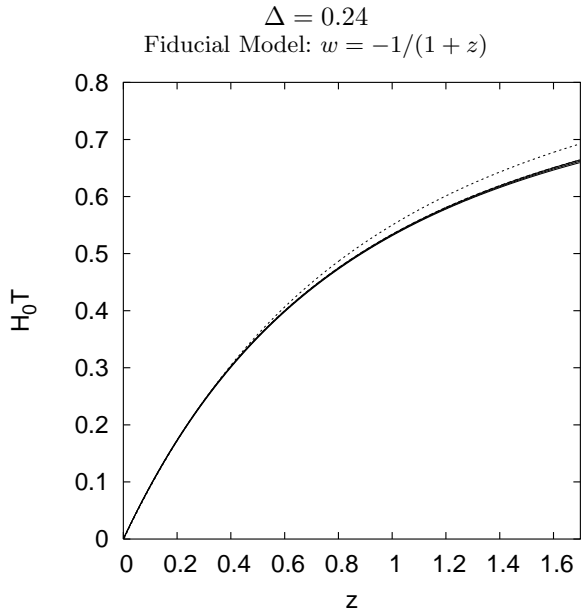,width=0.6\textwidth,angle=-90}
\vspace{-0.6cm}
\end{center}
\caption{\small
The smoothing scheme of equation~(\ref{eq:bg}) is used to determine
the look-back time of the universe, $T(z)=t(0)-t(z)$, from $1000$
realizations of the SNAP dataset for a $w=-1/(1+z) $ `metamorphosis'
model. The smoothing scale is $\Delta=0.24$. The solid lines show the
mean look-back time and the $1\sigma$ limits around it. The look-back
time for the fiducial model matches exactly with the mean for the
smoothing scheme. The dotted line shows the $\Lambda$CDM model.}
\label{fig:age}
\end{figure}

\noindent{\bf Age of the Universe}

We may also use this smoothing scheme to calculate other cosmological
parameters of interest such as the age of the universe at a redshift $z$:
\beq
t(z)=H_{0}^{-1}\int_{z}^{\infty} \frac{dz'}{(1+z') H(z')} \,\,.
\eeq
In this case, since data is available only upto redshifts of $z\simeq
1.7$, it will not be possible to calculate the age of the universe.
Instead, we calculate the look-back time at each redshift--
\beq
T(z)=t(z=0)-t(z)=H_{0}^{-1}\int_{0}^{z} \frac{dz'}{(1+z') H(z')} \,\,.
\eeq

Figure~\ref{fig:age} shows the reconstructed $T(z)$ with $1\sigma$
errors for the $w=-1/(1+z)$ `metamorphosis' model using the SNAP
distribution. For this model the current age of the universe is about
$13$ Gyrs and the look-back time at $z\simeq 1.7$ is about $9$ Gyrs
for a Hubble parameter of $H_0=70$ km/s/Mpc. We see that the look-back
time is reconstructed extremely accurately. Using this method we may
predict this parameter with a high degree of success and distinguish
between the fiducial look-back time and that for $\Lambda$CDM even at the
$10\sigma$ confidence level. Indeed any cosmological parameter which
can be obtained by integrating the Hubble parameter will be
reconstructed without problem, since integrating involves a further
smoothing of the results.

Looking at these results, we draw the conclusion that the method of
smoothing supernova data can be expected to work quite well for future
SNAP data as far as the Hubble parameter is concerned. Using this
method, we may reconstruct the Hubble parameter and therefore the
expansion history of the universe accurately.  We find that the method
is very efficient in reproducing $H(z)$ to an accuracy of $\lleq 2 \%$
within the redshift interval $0<z<1$, and to $\lleq 4 \%$ at
$z\simeq1.7$, as demonstrated in figure~\ref{fig:w=var,D=0.24,SNAP}.
Furthermore, using the Hubble parameter, one may expect to
discriminate between different families of models such as the
metamorphosis model $w=-1/(1+z)$ and $\Lambda$CDM. This method also
reproduces very accurately the look-back time for a given model, as
seen in fig~\ref{fig:age}. It reconstructs the look-back time to an
accuracy of $\lleq 0.2 \%$ at $z \simeq 1.7$.

\section{Reducing noise through Double Smoothing}\label{sec:double}

As we saw in the preceding section, the method of smoothing supernova
data to extract information on cosmological parameters works very well
if we employ the first derivative of the data to reconstruct the
Hubble parameter. It also works reasonably for the second derivative, which
is used to determine $w(z)$, but the errors on $w(z)$ are somewhat
large.  In this section, we examine a possible way in which the
equation of state may be extracted from the data to give slightly better
results.

\begin{figure*}
\begin{center}
\centerline{\mbox{{\large $\Delta=0.24$, Double Smoothing}}}
\centerline{\mbox{{\normalsize Fiducial Model: $w=-1/(1+z)$}}}
\psfig{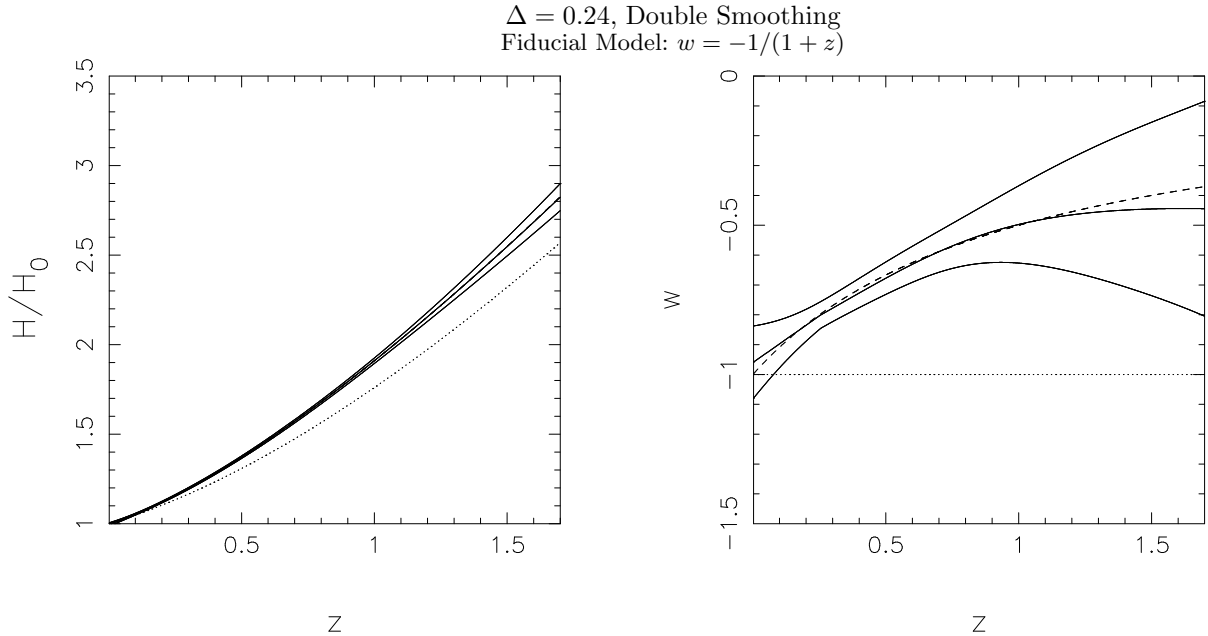}
\vspace{-0.6cm}
\end{center}
\caption{\small
The double smoothing scheme of equations~(\ref{eq:bg})
and~(\ref{eq:bg,double}) has been used to obtain $H(z)$ and $w(z)$
from $1000$ realizations of the SNAP dataset.  The smoothing scale is
$\Delta=0.24$.  The dashed line in each panel represents the fiducial
$w=-1/(1+z)$ `metamorphosis' model while the solid lines represent the
mean and $1\sigma$ limits around it.  The dotted line in both panels
is $\Lambda$CDM. In the left panel $H(z)$ for the fiducial model matches exactly with the mean for the smoothing scheme.}
\label{fig:w=var,D=0.24,SNAP,double}
\end{figure*}

\begin{figure*}
\begin{center}
\centerline{\mbox{{\large $\Delta=0.24$, Double Smoothing}}}
\centerline{\mbox{{\normalsize Fiducial Model: $w=-1$}}}
\psfig{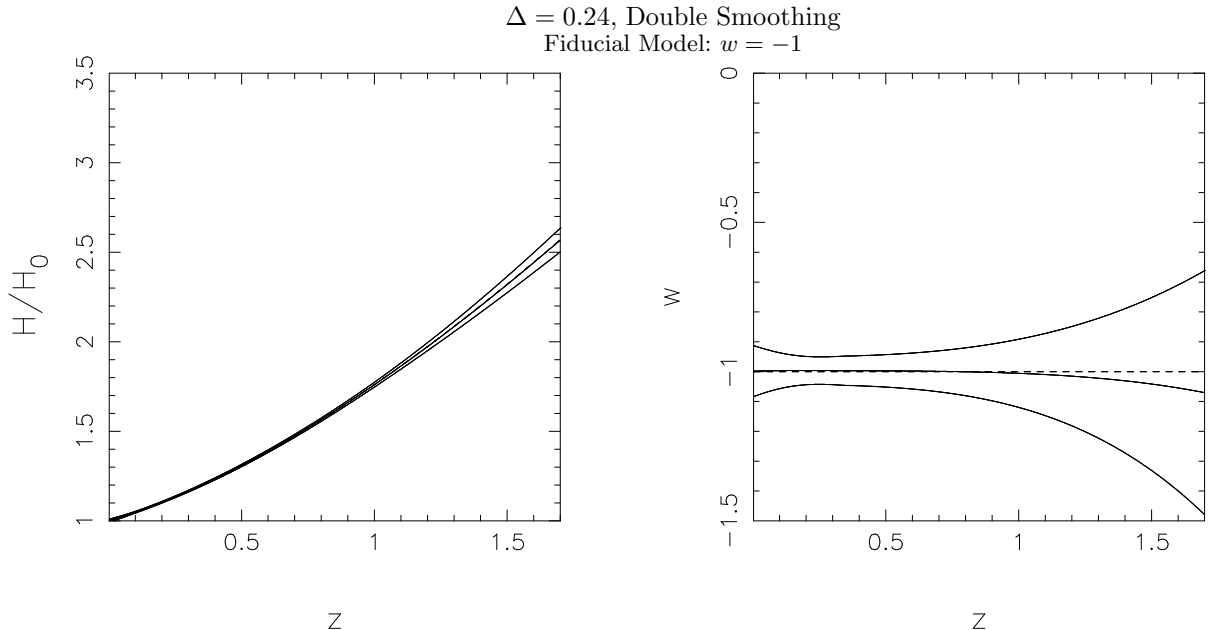}
\vspace{-0.6cm}
\end{center}
\caption{\small
The double smoothing scheme of equations~(\ref{eq:bg})
and~(\ref{eq:bg,double}) has been used to obtain $H(z)$ and $w(z)$
from $1000$ realizations of the SNAP dataset.  The smoothing scale is
$\Delta=0.24$.  The dashed line in each panel represents the fiducial
$\Lambda$CDM model with $w=-1$ while the solid lines represent the
mean and $1\sigma$ limits around it. In the left panel $H(z)$ for the fiducial model matches exactly with the mean for the smoothing scheme.}
\label{fig:w=-1.0,D=0.24,SNAP,double}
\end{figure*}

\begin{figure*}
\begin{center}
\centerline{\mbox{{\large $\Delta=0.24$, Double Smoothing}}}
\centerline{\mbox{{\normalsize Fiducial Model: $w=-0.5$}}}
\psfig{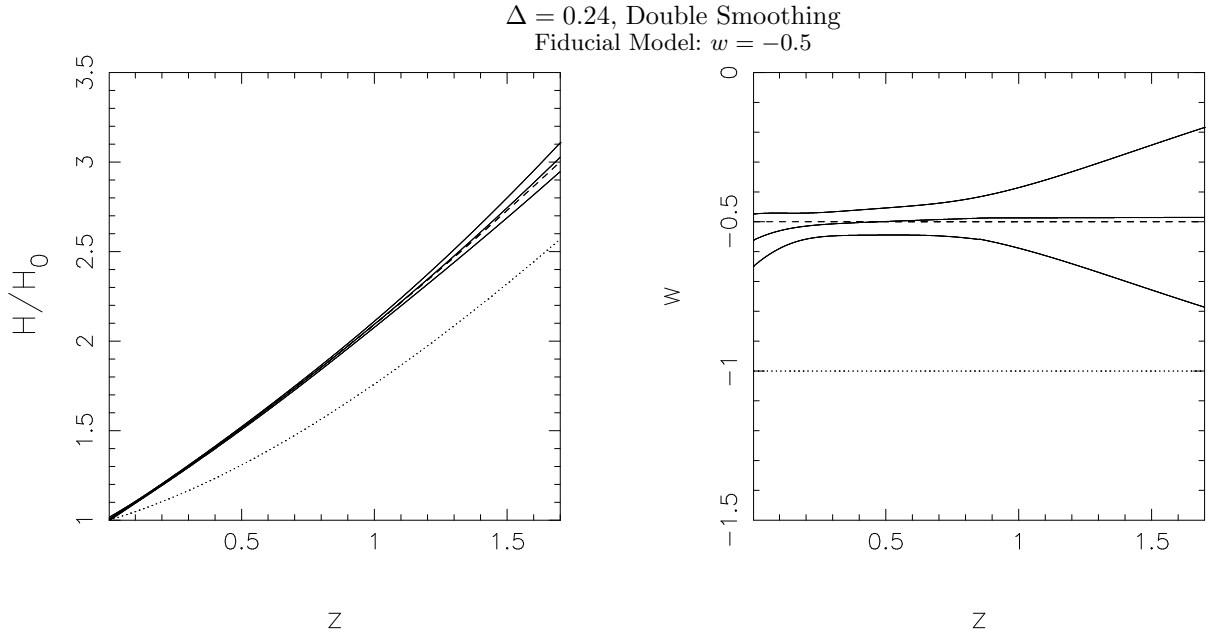}
\vspace{-0.6cm}
\end{center}
\caption{\small
The double smoothing scheme of equations~(\ref{eq:bg})
and~(\ref{eq:bg,double}) has been used to obtain $H(z)$ and $w(z)$
from $1000$ realizations of the SNAP dataset.  The smoothing scale is
$\Delta=0.24$.  The dashed line in each panel represents the fiducial
quiessence model with $w=-0.5$ while the solid lines represent the
mean and $1\sigma$ limits around it.  The dotted line is $\Lambda$CDM.  }
\label{fig:w=-0.5,D=0.24,SNAP,double}
\end{figure*}

The noise in each parameter translates into larger noise levels on its
successive derivatives. We have seen earlier that, using the smoothing
scheme~(\ref{eq:bg}), one can obtain $H(z)$ from the smoothed $d_L(z)$
fairly successfully. However, small noises in $H(z)$ propagate into
larger noises in $w(z)$.  Therefore, it is logical to assume that if
$H(z)$ were smoother, the resultant $w(z)$ might also have smaller
errors. So, we attempt to smooth $H(z)$ a second time after obtaining
it from $d_L(z)$. The procedure in this method is as follows -- first,
we smooth noisy data $\ln d_L(z)$ to obtain $\ln d_L(z)^{\rm s}$ using
equation~(\ref{eq:bg}).  We differentiate this to find $H(z)^{\rm s}$
using equation~(\ref{eq:hubb}). We then further smooth this Hubble
parameter by using the same smoothing scheme at the new redshifts
\ber\label{eq:bg,double}
H(z,\Delta)^{\rm s_2}=H(z)^g+N(z) \sum_i
\left [ H(z_i)^{\rm s}- H(z_i)^g \right] &&\nonumber\\
{\large \times} \ {\rm exp} \left [- \frac{\ln^2 \left ( \frac{1+z_i}{1+z} \right ) }{2 \Delta^2}
\right ],&&\\
N(z)^{-1}=\sum_i {\rm exp} \left [- \frac{\ln^2
\left ( \frac{1+z_i}{1+z} \right ) }{2 \Delta^2} \right ]. \hspace{1.7cm}&&\nonumber
\eer
We then use this $H(z,\Delta)^{\rm s_2}$ to obtain $w(z)$ using
equation~(\ref{eq:w}). This has the advantage of making $w(z)$ less
noisy than before, while using the same number of parameters. However,
repeated smoothing can also result in the loss of information.

The result for the SNAP distribution using this double smoothing
scheme for the $w=-1/(1+z)$ model is shown in
figure~\ref{fig:w=var,D=0.24,SNAP,double}.  We use $\Delta=0.24$ for
smoothing both $\ln d_L(z)$ and $H(z)$.  Comparing with
figure~\ref{fig:w=var,D=0.24,SNAP}, we find that there is an
improvement in the reconstruction of $H(z)$ as well as $w(z)$. Thus,
errors on the Hubble parameter decrease slightly and errors on $w(z)$
also become somewhat smaller.

We now explore this scheme further for other models of dark energy. We
first consider a $w=-1$ $\Lambda$CDM model. In
figure~\ref{fig:w=-1.0,D=0.24,SNAP,double}, we show the results for
this model. We find that the Hubble parameter accurately reconstructed
and even $w$ is well reconstructed, with a little bias at high
redshift. The next model we reconstruct is a $w=-0.5$ quiessence
model. The results for double smoothing are shown in
fig~\ref{fig:w=-0.5,D=0.24,SNAP,double}. There is a little bias for
this model at the low redshifts, although it is still well within the
error bars.

We note that in all three cases, a slight bias is noticeable at low or
high redshifts. This is primarily due to edge effects-- since at low
(high) redshift, any particular point will have less (more) number of
supernovae to the left than to the right. Even by estimating the guess
model through an iterative process, it is difficult to completely get
rid of this effect. In order to get rid of this effect, we would
require to use much larger number of iterations for the guess model,
but this would result in very large errors on the parameters. However,
this bias is so small as to be negligible and cannot affect the
results in any way.

\begin{table*}
\centering \caption{
The reconstructed $w${\tt -probe}
($\bar{w}$), determined using ~Eq.~(\ref{eq:weight2})
(and its $1\sigma$ error) is listed for 1000 realizations of SNAP data.
Three fiducial models are used : $w=-1/(1+z)$, $w=-1$ ($\Lambda$CDM)
and $w=-0.5$.  We deploy the method of double smoothing with
$\Delta=0.24$ to determine $\bar{w}$.  }
\label{tab:weight} \footnotesize
\begin{center}
\begin{tabular}{c|cc|cc|cc}
\hline
&\multicolumn{2}{|c|}{$w=-1/(1+z)$}&\multicolumn{2}{|c|}{$w=-1$}&\multicolumn{2}{|c|}{$w=-0.5$}\\
$\Delta z$&$\bar{w}$&$\bar{w}_{\rm exact}$&$\bar{w}$&$\bar{w}_{\rm exact}$&$\bar{w}$&$\bar{w}_{\rm exact}$\\
\hline
$0-0.414$&$-0.839 \pm 0.019$&$-0.845$&$-1.001 \pm 0.017$&$-1.0$&$-0.489 \pm 0.025$&$-0.5$\\
&&&\\
$0.414-1$&$-0.595 \pm 0.033$&$-0.598$&$-1.009 \pm 0.038$&$-1.0$&$-0.506 \pm 0.039$&$-0.5$\\
&&&\\
$1-1.7$&$-0.471 \pm 0.069$&$-0.432$&$-1.017 \pm 0.087$&$-1.0$&$-0.493 \pm 0.075$&$-0.5$ \\
\hline
\end{tabular}
\end{center}
\end{table*}

Looking at these three figures, we can draw the following conclusions.
The Hubble parameter is quite well reconstructed by the method of
double smoothing in all three cases while the errors on the equation
of state also decrease.  At low and high redshifts, a very slight bias
persists. Despite this, the equation of state is reconstructed quite
accurately.  Also, since the average error in $w(z)$ is somewhat less
than that in the single smoothing scheme
(figure~\ref{fig:w=var,D=0.24,SNAP}), the equation of state may be
used with better success in discriminating different models of dark
energy using the double smoothing procedure.

\section{The \lowercase {$w$}-probe}\label{sec:avgw}

In this section we explore the possibility of extracting information
about the equation of state from the reconstructed Hubble parameter by
considering a weighted average of the equation of state, which we call
the $w${\tt -probe}. An important advantage of this approach is that there
is no need to go to the second derivative of the luminosity distance
for information on the equation of state. Instead, we consider the
weighted average of the equation of state \citep{alam03b}
\beq
\label{eq:weight1}
1+\bar{w}= \frac {1}{\delta \ln(1+z)}\int \left(1+w(z)\right)\frac{dz}{1+z}\,\,,
\eeq
which can be directly expressed in terms of the difference in dark
energy density $\tilde {\rho}_{DE}=\rho_{DE}/\rho_{0c}$ (where
$\rho_{0c}=3H^2_0/8\pi G$) over a range of redshift as
\ber\label{eq:weight2} 
1+\bar{w}(z_1,z_2)\hspace{-0.4cm}&=&\hspace{-0.3cm}\frac{1}{3} \frac{\delta \ \ln \tilde{\rho}_{DE}}
{\delta \ \ln (1+z)}~~\nonumber\\
&=& \hspace{-0.4cm}\frac{1}{3}{\rm ln}\left[\frac{H^2(z_1)-\Omega_{0\rm m} (1+z_1)^3}
{H^2(z_2)-\Omega_{0\rm m} (1+z_2)^3}\right] {\Big /} {\rm ln}
\left(\frac{1+z_1}{1+z_2}\right)\nonumber\\
\eer
where $\delta$ denotes the total change of a variable between
integration limits.  Thus, even if the equation of state is noisy, the
$\bar{w}$ parameter may be obtained accurately provided the Hubble
parameter is well constructed. 

The parameter $\bar{w}$ has the interesting property that for the
concordance $\Lambda$CDM model, it equals $-1$ in all redshift ranges
while for other models of dark energy it is non-zero. 
For (non-$\Lambda$CDM) models with
constant equation of state, this parameter is a constant (but not
equal to $-1$), while for models with variable equation of state, it
varies with redshift. The fact that $\Lambda$CDM is a fixed
point for this quantity may be utilized to differentiate between the
concordance $\Lambda$CDM model and other models of dark energy.
Therefore the parameter $\bar{w}$ may be used as a new diagnostic
of dark energy which acts as a discriminator between $\Lambda$CDM and
other models of dark energy. We call this diagnostic the 
  $w${\tt -probe}.

We now calculate the $w${\tt -probe} for the three models described
above using the method of double smoothing.  In table \ref{tab:weight},
we show the values of $\bar{w}$ obtained in different redshift ranges
after applying double smoothing on SNAP-like data. The ranges of
integration are taken to be approximately equally spaced in ${\rm ln}
(1+z)$. Two points of interest should be noted here: a) $\bar{w}$
is very close to $\bar{w}_{\rm exact}$ at all redshifts for all
three models of dark energy; and b) as expected, this parameter is
good at distinguishing between $\Lambda$CDM and other dark energy
models.

\begin{figure}
\begin{center}
\vspace{0.3cm}
\psfig{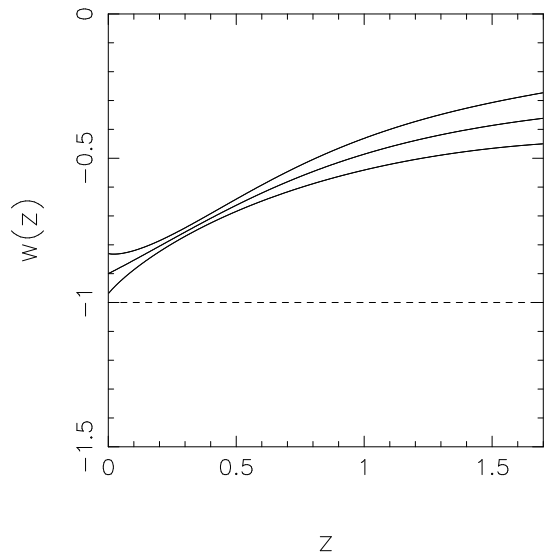}
\vspace{-0.6cm}
\end{center}
\caption{\small
The reconstructed equation of state $w(z)$ 
is shown for $1000$ realizations of
a $\omt=0.3, w=-1$ $\Lambda$CDM model. We assume an incorrect value for the matter
density, $\omt=0.2$ in the reconstruction exercise performed using (\ref{eq:w})
and the ansatz (\ref{eq:h_poly}). This is done to
study the effect of the observed uncertainty in $\omt$ on the
equation of state.  
The dashed line represents the fiducial
$\Lambda$CDM model with $w=-1.0$.  The solid lines 
represent the mean $w(z)$ and the $1\sigma$ limits around
it. }
\label{fig:h_exp_lcdm_om=0.2}
\end{figure}

In the above analysis, we have assumed that the matter density is
known exactly, $\omt=0.3$. Studies of large scale structure and CMB
have resulted in very tight bounds on the matter density, but still
some uncertainty remains regarding its true value.  As noted in
\cite{maor02}, a small uncertainty in the value of $\omt$ may affect
the reconstruction exercise quite dramatically. The Hubble parameter
is not affected to a very high degree by the value of matter density,
because it can be calculated directly as the first derivative of the
luminosity distance, which is the measured quantity.  However, when
calculating the equation of state of dark energy, the value of $\omt$
appears in the denominator of the expression~(\ref{eq:w}), hence any
uncertainty in $\omt$ is bound to affect the reconstructed $w(z)$. 
This is illustrated in figure \ref{fig:h_exp_lcdm_om=0.2}
where the equation of state is 
reconstructed for a fiducial $\Lambda$CDM model with $\omt=0.3$
by assuming that 
the matter density has been {\em incorrectly} determined to be
$\omt=0.2$, and this incorrect value is then used to determine
$w(z)$. 
For this purpose we consider a
complementary reconstruction ansatz that uses a polynomial fit to the
dark energy density \citep{rs1}
\beq\label{eq:h_poly}
H^2(z)=H_0^2 [\omt (1+z)^3+A_0+A_1(1+z)+A_2(1+z)^2]\,\,,
\eeq
where $A_0=1-\omt-A_1-A_2$ for a flat universe. This ansatz is known
to give accurate results for the dark energy density
\citep{alam03b,alam04a}.
Figure \ref{fig:h_exp_lcdm_om=0.2} clearly shows that 
an incorrect value of $\omt$ gives rise to an erroneously evolving
equation of state of dark energy whereas, in fact, the correct EOS
remains fixed at $w=-1$ and does not evolve.

One
of the main results of this paper is that, although the equation of
state $w(z)$ may be reconstructed badly if $\omt$ is not known
accurately, the uncertainty in $\omt$ does not have such a strong
effect on the reconstruction of the $w${\tt -probe} $\bar{w}$.  This
is because $\bar{w}$ in equation~(\ref{eq:weight2}) is a difference of
two terms, both involving $\omt$. As a result, uncertainty in $\omt$
does not affect $\bar{w}$ as much as it affects $w(z)$.  Therefore,
even when $\omt$ is not known to a high degree of accuracy, the
$w${\tt -probe} may still be reconstructed fairly accurately.

\begin{figure*}
\begin{center}
\psfig{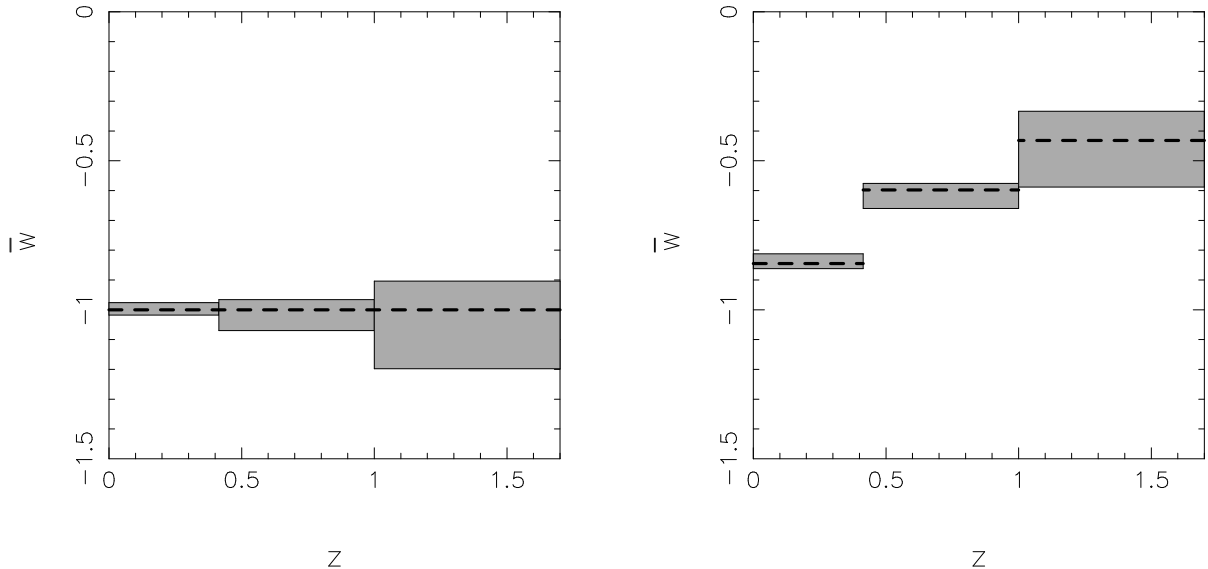}
\vspace{-0.6cm}
\end{center}
\caption{\small
The $w$-probe is reconstructed for the unevolving $\Lambda$CDM model
with $w=-1$ (left panel) and an evolving DE model with $w=-1/(1+z)$
(right panel). $1000$ realizations of SNAP-like data have been used.
The thick dashed line in both panels indicates the exact value of
$\bar{w}$ for the fiducial model, the dark grey boxes in each panel
indicate the $1\sigma$ confidence levels on $\bar{w}$ reconstructed
for the two models using the double smoothing scheme with
$\Delta=0.24$ and marginalising over $\omt=0.3 \pm 0.07$. This figure
illustrates that the $w$-probe works remarkably well for both
$\Lambda$CDM (left panel) and for evolving DE (right panel). The
details for this figure are given in Table~\ref{tab:margin}.}
\label{fig:margin}
\end{figure*}

We now demonstrate this by showing the results obtained using our
smoothing scheme after marginalising over the matter density. We
simulate SNAP like data for two models: (a) $\Lambda$CDM and (b) a
$w=-1/(1+z)$ `metamorphosis' model. When applying the smoothing
scheme, we assume that $\omt$ follows a Gaussian probability
distribution with mean $\overline{\omt}=0.3$ and variance
$\sigma=0.07$ (the error being commensurate to that expected from the
current CMB and Large Scale Structure data \citep{2dF}).  In
figure~\ref{fig:margin} and table~\ref{tab:margin}, we show the
results for the $w${\tt -probe} calculated for the two models.  We
find that the $w${\tt -probe} ($\bar{w}$) is determined to a high
degree of accuracy for both the models {\em even when we marginalize
  over $\omt$} ! The value of $\bar{w}$ for the $\Lambda$CDM model is
approximately equal to $-1$, while that for the metamorphosis model
shows clear signature of evolution. Thus, even if the matter density
of the universe is known uncertainly, this uncertainty does not affect
the accuracy of the reconstructed $w${\tt -probe} significantly. This
is a powerful result since it indicates that unlike the equation of
state, the $w${\tt -probe} is not overtly sensitive to the value of
$\omt$ for SNAP-quality data.

\begin{table}
\centering \caption{
The reconstructed $w${\tt -probe}
$\bar{w}$~(Eq.~(\ref{eq:weight2})) over specified redshift ranges
(and its $1\sigma$ error) is shown for 1000 realizations of SNAP data.
Two fiducial models are used : the $w=-1/(1+z)$ `metamorphosis' model
and $w=-1$ ($\Lambda$CDM). We deploy the method of double smoothing
with $\Delta=0.24$ and marginalize over $\omt=0.3 \pm 0.07$.}
\label{tab:margin} \footnotesize
\begin{center}
\begin{tabular}{c|cc|cc}
\hline
&\multicolumn{2}{|c|}{$w=-1/(1+z)$}&\multicolumn{2}{|c|}{$w=-1$}\\
\hline
$\Delta z$&$\bar{w}$&$\bar{w}_{\rm exact}$&$\bar{w}$&$\bar{w}_{\rm exact}$\\
\hline
$0-0.414$&$-0.837 \pm 0.025$&$-0.845$&$-1.003 \pm 0.021$&$-1.0$\\
&&&\\
$0.414-1$&$-0.618 \pm 0.042$&$-0.598$&$-1.018 \pm 0.052$&$-1.0$\\
&&&\\
$1-1.7$&$-0.461 \pm 0.127$&$-0.432$&$-1.051 \pm 0.147$&$-1.0$ \\
\hline
\end{tabular}
\end{center}
\end{table}

From the above results, we see that the $w${\tt -probe} is very
effective as a diagnostic of dark energy, especially in
differentiating between $\Lambda$CDM and other models of dark energy.
We summarise some important properties of the $w${\tt -probe} below:

\begin{enumerate}
  
\item $\bar{w}(z_1,z_2)$ is determined from the {\em first} derivative 
of the luminosity distance. Its reconstructed value is therefore less
noisy than the equation of state $w(z)$ (which is determined after
differentiating $d_L(z)$ twice; compare (\ref{eq:w}) and (\ref{eq:weight2}) ).

\item $\bar{w}(z_1,z_2) = -1$ {\em uniquely} for concordance cosmology
($\Lambda$CDM). For all other dark energy models $\bar{w} \neq
-1$.  This remains true when $\bar{w}$ is marginalized over
$\omt$.

\item $\bar{w}$ is robust to small uncertainties in the value of the matter
density. As we saw earlier this uncertainty can induce large errors in
determinations of the cosmic equation of state $w(z)$, see also
\cite{maor02}.  The weak dependence of $\bar{w}$ on the
value of $\omt$ in the range currently favored by observations $0.2
\leq \Omega_{0m} \leq 0.4$ implies that the $w${\tt -probe} can cope very
effectively with the existing uncertainty in the value of the matter
density for SNAP-quality data.
Furthermore, since $\bar{w}$ is constructed
directly from $\rho_{\rm DE}$, any method which determines either the
dark energy density or the Hubble parameter from observations can be
used to also determine $\bar{w}$.  Note that several excellent methods
for determining $\rho_{\rm DE}$ and $H(z)$ have been suggested in the
literature \citep{daly03,daly04,alam03b,alam04a,wangm,wangteg05}, and
any of these could be used to great advantage in determining the
$w${\tt -probe}.

\end{enumerate}

Thus we expect that the $w${\tt -probe} may be used as a handy
diagnostic for dark energy, especially in discriminating between
$\Lambda$CDM and other models of dark energy, for SNAP like datasets.
Its efficacy lies in the fact that it is not very sensitive to both the
value of the present matter density and also the reconstruction method
used.

\section{Cosmological Reconstruction applied to other physical models 
of Dark Energy}\label{sec:compare}

In this section we draw the readers attention to the dangers
encountered during cosmological reconstruction of atypical dark energy
models. There are currently two plausible ways of making the expansion
of the universe accelerate at late times. The first approach depends
on changing the matter sector of the Einstein equations. Examples of
this approach are the quintessence fields. A completely different
approach has shown that it is possible to obtain an accelerating
universe through modifying the gravity sector (see, for instance,
\cite{DDG,cardass,ss02,carroll03,capozziello03,nojiri03,dolgov03,sahni05}
and references therein). In these models, dark energy should not be
treated as a fluid or a field. Instead, it may be better dubbed as 
'geometric dark energy'. Indeed the DGP model can cause the universe to 
accelerate even in the absence of a physical dark energy component.
As pointed out in \cite{alam03a,sahni05}, the equation of state is not
a fundamental quantity for geometric dark energy. E.g., using
$w(z)$ in the reconstruction of such models may result in very strange
results, including, for instance, singularities in the equation of
state.  \footnote{A very simple model which has a well-behaved $q(z)$
  but singular $w(z)$ is a model which has, in addition to the
  cosmological constant, a second dark energy component disguised as a
  spatial curvature term-- $H^2(z)/H_0^2=\omt (1+z)^3+\Omega_X (1+z)^2
  +\Omega_{\Lambda}$. If we assume that
  $\omt=0.3,\Omega_X=-0.05,\Omega_{\Lambda}=0.75$, then $w(z)$ becomes
  singular when $\Omega_X (1+z)^2 +\Omega_{\Lambda}=0$, i.e., at 
  $z \simeq 2.8$.
  Although this property of $w(z)$ can be easily understood physically
  and rests in the fact that it is an `effective' equation of
  state for the combination of DE fluids, nevertheless any reasonable
  parameterization of $w(z)$ will clearly experience difficulty in
  reproducing this behavior. An effective equation of state with a
  similar `pole-like' divergence is frequently encountered in
  braneworld models of dark energy \citep{ss02,loiter} as well as
  in holographic models \citep{linder04}. }

\begin{figure}
\begin{center}
\centerline{\mbox{{\large $\Delta=0.24$, Double Smoothing}}}
\centerline{\mbox{{\normalsize Fiducial Model: Braneworld : $\omt=0.3,\Omega_l=1,\Omega_{\l_b}=0$}}}
\vspace{0.2cm}
\psfig{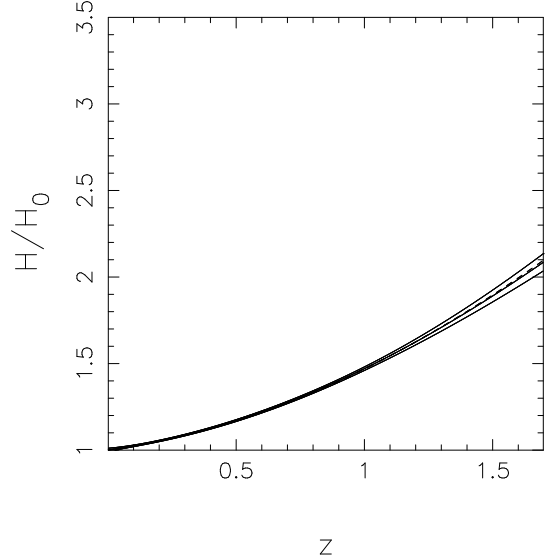}
\vspace{-0.6cm}
\end{center}
\caption{\small
The reconstructed Hubble parameter $H(z)$ for the braneworld
model~(\ref{eq:hubble_brane}) for $1000$ realizations of the SNAP
dataset. We use the double smoothing scheme of equations~(\ref{eq:bg})
and~(\ref{eq:bg,double}) with $\Delta =0.24$.  The dashed line
represents the fiducial $\omt=0.3,\Omega_l=1,\Omega_{\l_b}=0$
braneworld model while the solid lines represent the mean and
$1\sigma$ limits around it.  We note that the Hubble parameter is very
well reconstructed for the braneworld model.  }
\label{fig:brane_h}
\end{figure}

As an example, we consider the braneworld dark energy model proposed
in \cite{ss02} described by the following set of equations for a flat
universe : 
\ber\label{eq:hubble_brane}
{H^2(z) \over H_0^2} &=& \omt (1\!+\!z)^3 + \Omega_\sigma +\\
&&2 \Omega_l - 2 \sqrt{\Omega_l}\, \sqrt{\omt (1\!+\!z )^3 + \Omega_\sigma + \Omega_l + \Omega_{\l_{\rm b}}} \nonumber\\
\Omega_{\sigma} &=& 1-\omt+2\sqrt{\Omega_l(1+\Omega_{\l_b})}\,\,, 
\eer
where the densities $\Omega$ are defined as : 
\beq \label{eq:omegas}
\omt = {\rho_{0m} \over 3 m^2 H_0^2} , \Omega_\sigma = {\sigma \over 3 m^2H_0^2} , \Omega_l = {1 \over l_c^2 H_0^2} , \Omega_{\l_{\rm b}} = - {\l_{\rm b} \over 6 H_0^2} \,\,, 
\eeq 
$l_c = m^2/M^3$ being a new length scale ($m$ and $M$ refer
respectively to the four and five dimensional Planck masses), $\l_{\rm
  b}$ the bulk cosmological constant and $\sigma$ the brane tension. In this section we have
  used $\hbar =c=1$.
On short length scales $r \ll l_c$ and at early times, one recovers
general relativity, whereas on large length scales $r \gg l_c$ and at
late times brane-related effects become important and may lead to the
acceleration of the universe. The `effective' equation of state for
this braneworld model is given by
\ber
&&\rho=\frac{3H^2}{8\pi G} (1-\Omega_{\rm m}(z)), \ p=\frac{H^2}{4\pi G} (q(z)-1/2)\\
&&w_{\rm eff}=\frac{p}{\rho}=\frac{q(z)-1/2}{3(1-\Omega_{\rm m}(z))}\,\,.  
\eer 
It is obvious that the effective equation of state in this braneworld
model may become singular if $\Omega_{\rm m}(z)\equiv \omt (1+z)^3
H_0^2/H^2(z)$ becomes unity. This
does not signal any inherent pathologies in the model however.  We
should remember that the acceleration of the universe in this model is
due to modification of the expansion of the universe at late times
due to extra-dimensional effects. Hence it is not very appropriate to
describe dark energy by an equation of state for such a model.  However, 
it would be
interesting to see if the singularity in the effective $w$ for this
model can be recovered by our smoothing method.

\begin{figure*}
\begin{center}
\centerline{\mbox{{\normalsize Fiducial Model: Braneworld : $\omt=0.3,\Omega_l=1,\Omega_{\l_b}=0$}}}
\vspace{0.2cm}
\psfig{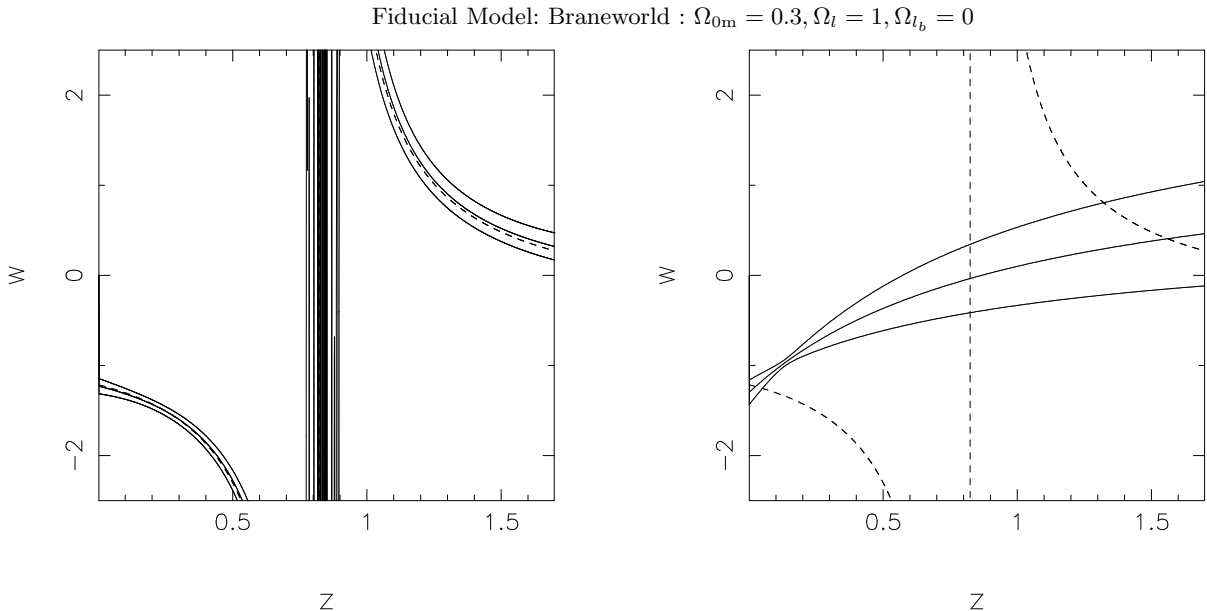}
\vspace{-0.6cm}
\end{center}
\caption{\small
Reconstructed equation of state for the braneworld
model~(\ref{eq:hubble_brane}) for $1000$ realizations of the SNAP
dataset. The left panel shows results for the double smoothing scheme
of equations~(\ref{eq:bg}) and~(\ref{eq:bg,double}) with $\Delta
=0.24$ while the right panel uses the ansatz~(\ref{eq:linder}) to reconstruct the
equation of state.  The dashed line in each panel represents the
fiducial $\omt=0.3,\Omega_l=1,\Omega_{\l_b}=0$ braneworld model while
the solid lines represent the mean and $1\sigma$ limits around it.  }
\label{fig:brane}
\end{figure*}

We attempt to reconstruct an $\omt=0.3,\Omega_l=1,\Omega_{\l_b}=0$
braneworld model which is a good fit to the current supernova data
\cite{snbrane}. We simulate data according to SNAP and obtain results
for the double smoothing method with $\Delta=0.24$.  In
figure~\ref{fig:brane_h}, we show the reconstructed Hubble parameter
for this reconstruction. We see that the Hubble parameter is very well
reconstructed and shows no pathological behavior. 

We now obtain the equation of state of dark energy for this model. For
this purpose, we also use an ansatz for the equation of state as
suggested by \cite{polar} and \cite{linder} (the CPL fit)
\beq\label{eq:linder}
w(z)=w_0+\frac{w_1 z}{1+z}\,\,.
\eeq

The results are shown in figure~\ref{fig:brane}. We find, as expected,
that it is impossible to catch the singularity in the equation of
state at $z\simeq0.8$ using an equation of state ansatz. Of course, one
may try and improve upon this somewhat dismal picture by introducing
fits with more free parameters.  However, it is well known that the
presence of more degrees of freedom in the fit leads to a larger
degeneracy (between parameters) and hence to larger errors of
reconstruction \citep{albrecht}.  In contrast to this approach, when
we reconstruct the equation of state using the smoothing scheme (which
does not presuppose any particular behavior of the equation of state),
the Hubble parameter is reconstructed very accurately and hence the
`effective' equation of state for this model is also reconstructed
well, as shown in figure~\ref{fig:brane}.  From this figure we see a
clear evidence of the singularity at $z \simeq 0.8$.  Thus to obtain
maximum information about the equation of state, especially in cases
where the dark energy model is very different from the typical
quintessence-like models, it may be better to reconstruct the Hubble
parameter or the dark energy density first.

Therefore, we find that the smoothing scheme, which performs reasonably
when reconstructing quintessence models of dark energy models, can be also
applied to models which show a
departure from general relativistic behavior at late times.
\footnote{Note, however, that most reconstruction methods including the
  present one may have problems in reproducing the rapidly
  oscillating equation of state predicted to arise in some models of
  dark energy \citep{oscillate}.}  This section illustrates the fact
that, in general, reconstructing $H(z)$ and its derivatives such as
the deceleration parameter $q(z)$ may be less fraught with difficulty
than a reconstruction of $w(z)$, which, being an effective equation of
state and not a fundamental physical quantity in some DE models, can
often show peculiar properties.

\section{Conclusion}

This paper presents a new approach to analyzing supernova data and
uses it to extract information about cosmological functions, such as
the expansion rate of the universe $H(z)$ and the equation of state of
dark energy $w(z)$.  In this approach, we deal with the data directly
and do not rely on a parametric functional form for fitting any of the
quantities $d_L(z), H(z)$ or $w(z)$. Therefore, we expect the results
obtained using this approach to be model independent. A Gaussian
kernel is used to smooth the data and to calculate cosmological
functions including $H(z)$ and $w(z)$. The smoothing scale used for
the kernel is related to the number of supernovae, errors of 
observations and derived errors of the parameters by a simple
formula, eq~(\ref{eq:tegmark}). For a given supernova distribution, the
smoothing scale determines both the errors on the parameters and the
bias of the results (see appendix~\ref{app:errors_bias}). $\Delta$
cannot be increased arbitrarily as this would diminish the reliability
of the results. We use a value of $\Delta$ which gives results which
have reasonably small bias as well as acceptable errors of $H(z)$ for
the SNAP quality data used in our analysis (see
section~\ref{sec:results}). As can be seen from eq~(\ref{eq:tegmark}),
when the data improves (i.e., the  number of data points increases and/or
measurement errors decrease), we expect that the same value of
$\Delta$ would result in smaller errors on $H(z)$.

We demonstrate that this method is likely to work very well with
future SNAP-like SNe data, especially in reconstructing the Hubble
parameter, which encodes the expansion history of the universe.
Moreover, our successful reconstruction of the Hubble parameter can
also be used to distinguish between cosmological models such as
$\Lambda$CDM and evolving dark energy.  The method can be further refined,
if one wishes to reconstruct the cosmic equation of state to greater
accuracy, by double smoothing the data-- smoothing the Hubble
parameter, after it has been derived from the smoothed luminosity
distance, so as to reduce noise in $w(z)$ (as in
section~\ref{sec:double}). The results obtained using the smoothing
scheme compare favorably to results obtained by other methods of
reconstruction.  Another quantity which may be reconstructed to great
accuracy is the look-back time of the universe.

An important result of this paper is the discovery that the
$w${\tt -probe} (originally proposed in \cite{alam03b}) provides us
with an excellent diagnostic of dark energy.  We summarize some of the
attractive features of this diagnostic below.

(a) The $w${\tt -probe} defined in Eqs. (\ref{eq:weight1}) and 
(\ref{eq:weight2}) is obtained from the luminosity distance by means
of a {\em single} differentiation.  Therefore, it avoids the pitfalls
of $w(z)$ which is obtained from the luminosity distance through a
double differentiation -- see Eq.(\ref{eq:w}), and hence is usually
accompanied by large errors (see also \cite{maor}).

(b) The $w${\tt -probe} is robust to small uncertainties in the value
of $\Omega_{0m}$. This attractive property allows us to get around 
observational uncertainties in the value of $\Omega_{0m}$ currently
known to an accuracy of about $30\%$. Indeed, when marginalized over
$\Omega_{0m}$, the $w${\tt -probe} can be used to great advantage to
distinguish between $\Lambda$CDM 
and other dark energy models for SNAP-quality data.

We therefore conclude that the proposed reconstruction method by
smoothing the supernova data appears to be sufficiently accurate and,
when applied to SNAP-type observations, should be able to distinguish
between evolving dark energy models and a cosmological constant. 

The method proposed by us can also be used for other forms of data which deliver the luminosity (or angular size) distance. 


\section{Acknowledgment}

We would like to thank G. Aldering for providing us with the
distribution of SNAP supernovae (table~\ref{tab:SNAP}). UA thanks the
CSIR for providing support for this work.  AAS was partially supported
by the Russian Foundation for Basic Research, grant 05-02-17450, and
by the Research Program `Astronomy' of the Russian Academy of
Sciences.

\appendix

\section{Smoothing errors and bias}\label{app:errors_bias}

In this section we explore the errors on the cosmological parameters
due to the smoothing scheme, as also the bias which enters the
results.

\subsection{Smoothing errors}

The smoothing scheme used in this paper is of the form :
\beq\label{eq:smooth}
y(z)^{s}=y_G(z)+\sum_{i=1}^N [y(z_i)-y_G(z_i)]S(z_i;\Delta)/\sum_{i=1}^N S(z_i;\Delta)\,\,,
\eeq
where the quantity $S(z_i;\Delta)$ represents the smoothing function
with a scale $\Delta$ and $y_G(z)$ is the subtracted guess model. The
quantity being smoothed (in this case $\ln d_L$) is represented by
$y$, while $y^{s}$ represents
the smoothed result.
Let the errors in the data at any redshift $z_i$ be given by
$\sigma_{y}(z_i)$ and the errors in the guess model be
$\sigma_{y_G}(z_i)$. If we look at the second term on the right hand
side of eq~(\ref{eq:smooth}), we see that the errors on this term
would be approximately given by the errors on $y$ weighted down by the
smoothing scale $\Delta$ and the number of data points $N$. Therefore
the error on the smoothed result is:
\beq
\sigma_{y_{s}}^2(z) \simeq \sigma_{y_G}^2(z)+\frac{\sigma_y^2(z)+\sigma_{y_G}^2(z)}{N \Delta}\,\,.
\eeq

We now consider the errors for an iterative method. The first guess is
an exact model, $\Lambda$CDM. Therefore the error on the result of
the first iteration is simply
\beq
\sigma_{y_1}^2(z) \simeq \frac{\sigma_y^2(z)}{N \Delta}\,\,.
\eeq
The next guess model is $y_1(z)$. Therefore the error on the result is
\beq
\sigma_{y_2}^2(z) \simeq \left( 2+ \frac{1}{N\Delta}\right)\frac{\sigma_y^2(z)}{N \Delta}\,\,.
\eeq

From this we can show that the errors on the result for the M-th iteration is :
\ber
\sigma_{y_M}^2(z) &\simeq& \left[1+\sum_{i=1}^{M-1}\left(1+\frac{1}{N\Delta}\right)^i\right]\frac{\sigma_y^2(z)}{N \Delta}  \nonumber\\
&\simeq& \left[M+ \frac{M(M-1)}{2 N\Delta}\right]\frac{\sigma_y^2(z)}{N \Delta}\,\,.
\eer
The second term on the right-hand side is small for a reasonable
number of iterations, since $N \simeq 2000$ and $\Delta > 0.01$
usually. Therefore we may approximate the errors on the log luminosity
distance after $M$ iterations for the guess model as
\beq
\sigma_{\ln d_L}^M(z)=\sqrt{M}\sigma_{\ln d_L}^0(z)\,\,,
\eeq
where $\sigma_{\ln d_L}^0(z)$ is the error for a simple smoothing
scheme where the data is smoothed without using a guess model.

\begin{figure*}
\centering
\begin{center}
\psfig{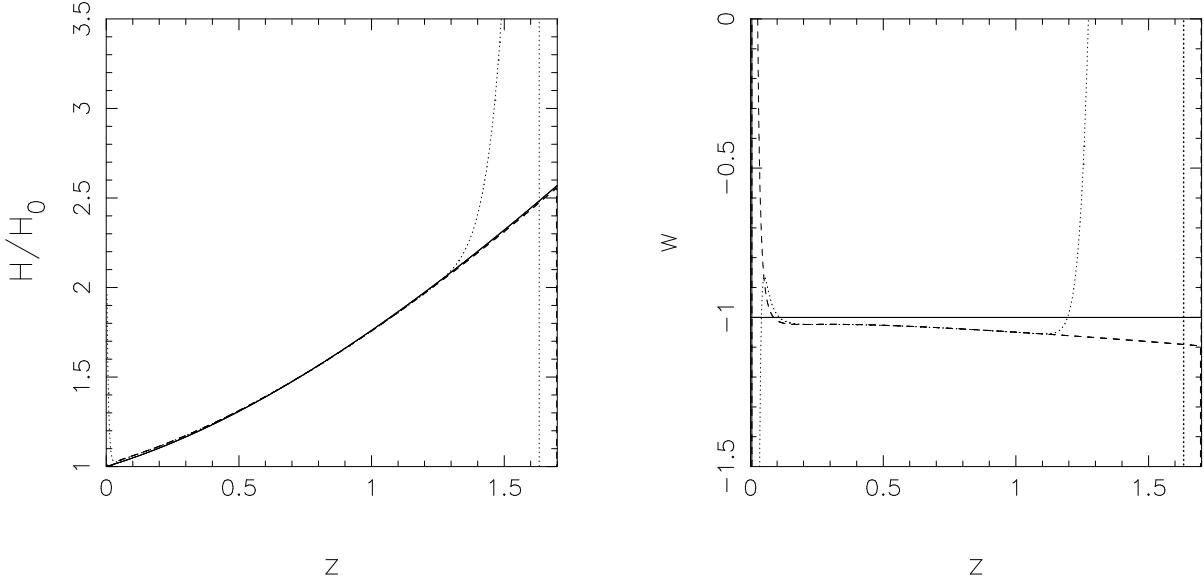}
\vspace{-0.5cm}
\end{center}
\caption{\small
Comparison of the reconstructed parameters obtained by using the
smoothing method (\ref{smooth_alexei}) with the variable
$\Delta(z)=0.2z/(1+z)^2$ for the $\Lambda CDM$ model (the
dotted line in each panel) with the analytical biased result given by
Eq. (\ref{eq:bias}) (dashed line).  The solid line represents the
fiducial $\Lambda CDM$ model.  Note the excellent agreement between
the analytical and numerical results in the redshift range $0.1 \lleq
z \lleq 1.1$.  }
\label{fig:bias}
\end{figure*}

\subsection{Smoothing Bias}

In any kind of a smoothing scheme for the luminosity distance, some
bias is introduced both in it and in derived quantities like $H(z)$
and $w(z)$. To illustrate the effect of this bias, we calculate it for
the simplest Gaussian smoothing scheme for $\ln d_L(z)$ with the width
$\Delta(z)\ll 1$ :
\ber
\label{smooth_alexei}
\ln d_L(z)^{\rm s} = N(z) \sum_{i=1}^{M}\ln d_L(z_i)
{\rm exp} \left [- \frac{(z-z_i)^2}{2\Delta^2} \right ]~,&&\\
N(z)^{-1}= \sum_{i=1}^{M} {\rm exp}
\left [- \frac{(z-z_i)^2}{2\Delta^2} \right ],\hspace{2.2cm}&&\,\,
\eer
where M is the total number of supernovae data points. The bias at
each redshift (${\cal B}(z) = \ln d_L(z)^{\rm s} - \ln d_L(z)$)
is the difference between the smoothed $\ln d_L(z)$ and the exact
value of $\ln d_L(z)$ :
\beq
{\cal B}(z)=  N(z)
\sum_{i=1}^{M} (\ln d_L(z_i)-\ln d_L(z))\ {\rm exp}
\left [- \frac{(z-z_i)^2}{2\Delta^2} \right ]\,\,.
\eeq
Expanding $\ln d_L(z_i)$ in terms of $\ln d_L(z)$ and its derivatives
by Taylor expansion, we get:
\ber
\hspace{-0.4cm} {\cal B}(z)=  N(z)
\sum_{i=1}^{M} \left[(\ln d_L(z))'(z_i-z)+(\ln d_L(z))^{\prime\prime} \frac{(z_i-z)^2}{2} \right ] &&\nonumber\\ 
{\large \times} \ {\rm exp} \left [- \frac{(z-z_i)^2}{2\Delta^2}
\right ],&&\,\,
\label{eq:expan}
\eer
where the prime denotes the derivative with respect to $z$ and we
neglect higher derivatives.  To see the effect of this bias at low and
high redshifts where the number of supernovae on both sides of each
$z$ are not equal, we rewrite Eq. (\ref{eq:expan}) in another way.
Let $\delta$ be the spacing between two neighboring data points, so
that $z=m \delta$. For $m<M/2$, we have:
\begin{eqnarray*}
{\cal B}(z)=  \hspace{7.8cm}&& \nonumber\\
N(z)
\sum_{i=1}^{2m} \left [(\ln d_L(z))^{\prime\prime}\frac{\delta^2(i-m)^2}{2} \right ]
{\rm exp} \left [- \frac{\delta^2(i-m)^2}{2\Delta^2} \right ] &&\nonumber\\
+ N(z) \sum_{i=2m+1}^{M} \left [(\ln d_L(z))^{\prime}
\delta(i-m)+(\ln d_L(z))^{\prime\prime}\frac{\delta^2(i-m)^2}{2} \right ]&&\nonumber\\
{\large \times} \ {\rm exp} \left [- \frac{\delta^2(i-m)^2}{2\Delta^2} \right ]
&&\,\,,
\end{eqnarray*}

and for $m>M/2$:
\begin{eqnarray*}
{\cal B}(z)= \hspace{7.8cm}&& \nonumber\\
N(z)
\sum_{i=2m-M+1}^{M} \left [(\ln d_L(z))^{\prime\prime}
\frac{\delta^2(i-m)^2}{2} \right ] {\rm exp}
\left [- \frac{\delta^2(i-m)^2}{2\Delta^2} \right ] &&\nonumber\\
+ N(z) \sum_{i=1}^{2m-M} \left [(\ln d_L(z))^{\prime}
\delta(i-m)+(\ln d_L(z))^{\prime\prime}\frac{\delta^2(i-m)^2}{2} \right ]&&\nonumber\\
{\large \times} \ {\rm exp} \left [- \frac{\delta^2(i-m)^2}{2\Delta^2} \right ]&&
\end{eqnarray*}
The first term in the above equations is the general bias of the
method, while the second term is the bias arising due to an asymmetric
number of data points around each supernova. For $m=M/2$, the number
of data points is the same from both sides and we have:
\begin{eqnarray*}
{\cal B}(z)=  N(z) \sum_{i=1}^{M}
\left [(\ln d_L(z))^{\prime\prime}\frac{\delta^2(i-m)^2}{2} \right ] {\rm exp}
\left [- \frac{\delta^2(i-m)^2}{2\Delta^2} \right ].\nonumber
\end{eqnarray*}
In the continuous limit where $x=i-m$ is assumed, we get:
\ber
{\cal B}(z)&=&  N(z)
\int [(\ln d_L(z))^{\prime\prime}\frac{\delta^2 x^2}{2}]
{\rm exp} \left [- \frac{\delta^2 x^2}{2\Delta^2} \right ] dx~, \\
N(z)&=& \int {\rm exp} \left [- \frac{\delta^2 x^2}
{2\Delta^2} \right ] dx \nonumber\,\,.
\eer
Therefore, the bias has the simple form
\beq
\label{eq:bias}
{\cal B}(z)= \frac {(\ln d_L(z))^{\prime\prime}\delta^2}{2\delta^2/\Delta^2} =
\frac{\Delta^2}{2}(\ln d_L(z))^{\prime\prime}\,\,.  
\eeq 
This is a good analytical approximation for the bias at redshifts in
the middle range, where we do not encounter the problem of data
asymmetry. To see the effect of this bias, let us assume that the real
model is the standard $\Lambda CDM$, add the bias term to this model
and then calculate the biased $H(z)$ and $w(z)$. The result from this
analytical calculation can be compared to the result of smoothing the
exact $\Lambda CDM$ model using our method.  The
figure~\ref{fig:bias} simply illustrates that the results obtained
using Gaussian smoothing and by the use of formula (\ref{eq:bias}) are
in good agreement in the middle range of redshifts. However, we do not
expect the formula (\ref{eq:bias}) to work properly at very low
($z<0.1$) and high ($z>1$) redshifts where the above mentioned
asymmetry of points adds a further bias.

Also, it appears that the smoothing bias has a tendency to decrease
$w(z)$ below its actual value in the middle range of $z$.  Thus,
$\Lambda CDM$ may appear to be a `phantom' ($w<-1$) if too large a
smoothing scale is chosen.

\section{Exploring smoothing with variable width $\Delta(z)$}\label{sec:var}

\begin{figure*}
\centering
\begin{center}
\centerline{\mbox{{\large $\Delta=1.2\,{\frac{z}{(1+z)^2}} $}}}
\vspace{0.2cm}
\centerline{\mbox{{\normalsize Fiducial Model: $w=-0.5$}}}
\psfig{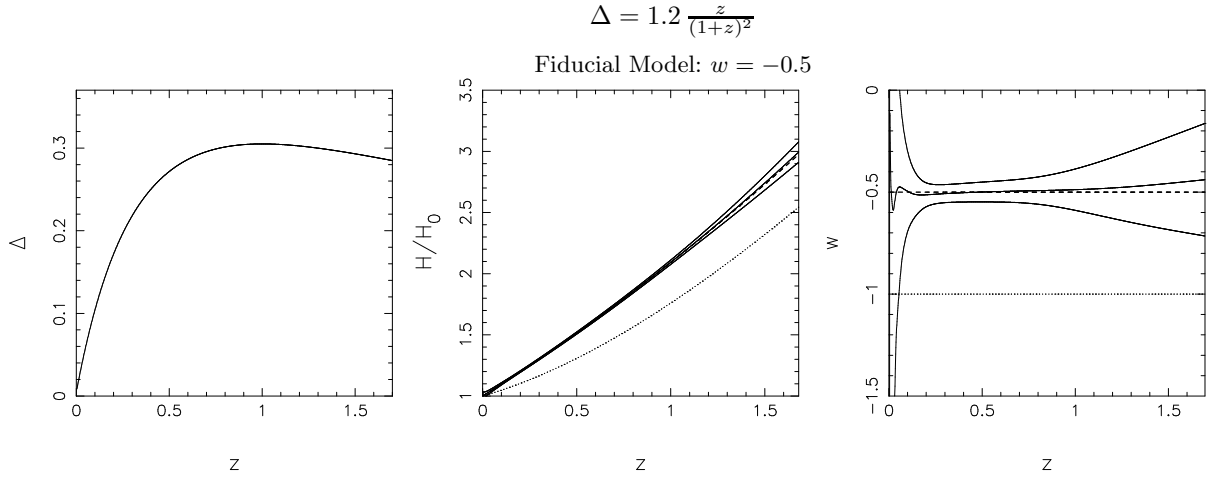}
\vspace{-0.6cm}
\end{center}
\caption{\small
The smoothing scheme of equation~(\ref{eq:bg}) is used with
$\Delta(z)=1.2\, z/(1+z)^2$ to obtain smoothed $H(z)$ and $w(z)$ from
$1000$ realizations of the SNAP dataset.  The panel (a) represents the
form of $\Delta(z)$ used, while panels (b) and (c) represent the
reconstructed $H(z)$ and $w(z)$.  The dashed line in panels (b) and
(c) represents the fiducial `metamorphosis' model with $w=-0.5$
while the solid lines represent the mean and $1\sigma$ limits around
it.  The dotted line is $\Lambda$CDM.  }
\label{fig:w=-0.5,alex,SNAP}
\end{figure*}

\begin{figure*}
\centering
\begin{center}
\centerline{\mbox{{\large $\Delta=0.36\,{\rm tanh}\,{\frac{0.23+z}{0.64}} $}}}
\vspace{0.2cm}
\centerline{\mbox{{\normalsize Fiducial Model: $w=-0.5$}}}
\psfig{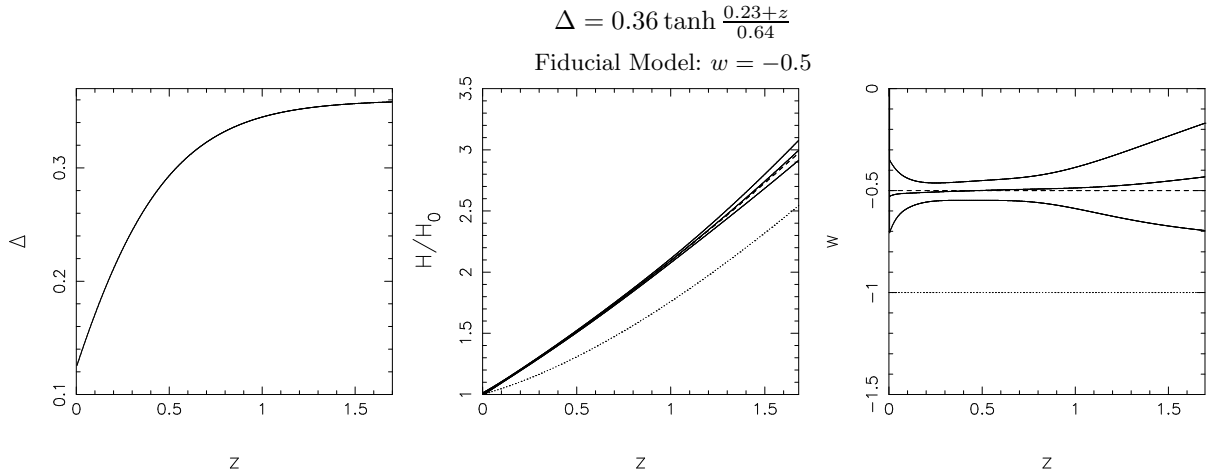}
\vspace{-0.6cm}
\end{center}
\caption{\small
The smoothing scheme of equation~(\ref{eq:bg}) is used with a tangent
hyperbolic form of variable $\Delta(z)$ to obtain smoothed $H(z)$ and
$w(z)$ from $1000$ realizations of the SNAP dataset.  The panel (a)
represents the form of $\Delta(z)$ used, while panels (b) and (c)
represent the reconstructed $H(z)$ and $w(z)$.  The dashed line in
panels (b) and (c) represents the fiducial $w=-0.5$
`metamorphosis' model while the solid lines represent the mean and
$1\sigma$ limits around it.  The dotted line is $\Lambda$CDM.  }
\label{fig:w=-0.5,D=tanh,SNAP}
\end{figure*}

In order to deal with the problem of data asymmetry and paucity at low
and high redshifts we may consider using a variable $\Delta(z)$. (i)
Low $z$ ($z_* \ll 1$) : in this case, there are many more supernovae
at $ z > z_*$ than there are at $ z < z_*$.  The error-bars are also
small in the low redshift region.  Therefore, a smaller value of
$\Delta$ appears to be more appropriate at low $z$.  (ii) High $z$
($z_* > 1$) : in this case, there is considerably more data at $ z <
z_*$ than at $ z > z_*$.  However, at high $z$ the errors are
considerably larger than at low $z$, which suggests that in order to
avoid a noisy result we must use a larger value of $\Delta$ in this
region. In this section, we investigate two different functional forms
of $\Delta(z)$ with the above properties and show how they result in the 
reconstruction of the equation of state.

\subsection{$\Delta(z)=\Delta_0 z/(1+z)^2$}\label{sec:alexei}

In section~\ref{sec:method} we mentioned that, for $|z-z_i|\ll 1$, the
exponent in Eq. (\ref{eq:bg}) reduces to the form
$-(z-z_i)^2/2\Delta^2(1+z)^2$ and the effective Gaussian smoothing
scale becomes $\Delta (1+z)$. So if we use a variable $\Delta(z) =
\Delta_0 z/(1+z)^2$ then the effective Gaussian smoothing scale
approaches a constant at large $z$ and tends to a small value at small
$z$. The results obtained using this method are shown in
figure~\ref{fig:w=-0.5,alex,SNAP} for SNAP data, using the model
$w=-0.5$.  We find that, the result for the Hubble parameter does
not change much. However, the equation of state is somewhat better
reconstructed, but noisier at low redshift because of the small width
of smoothing.

\subsection{tan-hyperbolic form of $\Delta(z)$}\label{sec:tanh}

Tangent hyperbolic form for $\Delta (z)$ is another form of the
variable $\Delta(z)$ which can simultaneously satisfy both the low and
high $z$ requirements. It has a small value at low redshifts and
a bigger value at the higher redshifts. An additional important
property of this function is that it changes smoothly from low to high
$z$, which translates into a smoother second derivative $w(z)$ -- see
(\ref{eq:bg}) - (\ref{eq:w}).

A drawback of this method is that the tangent hyperbolic function
introduces a number of free parameters into the problem. However the
role of these parameters can be understood as follows. The tangent
hyperbolic function can be written in the general form
\beq
\label{eq:tanh}
\Delta (z)\, = \,a\,{\rm tanh}\,{\frac{b+z}{c}}~.
\eeq
As we saw earlier, if $\Delta$ is held constant, then optimal results
are obtained for $\Delta_0 \,=\,0.24$ in (\ref{eq:bg}) when we use
bootstrap iterative process. We therefore determine $a,b$ and $c$ in
(\ref{eq:tanh}) so that $\Delta(z) \approx \frac{1}{2} \Delta_0$ at $z
\approx 0$, and $\Delta(z) \approx \frac{3}{2} \Delta_0$ at $z \approx
1.7$; consequently
\beq
\Delta (z)\, =
\,0.36\,{\rm tanh}\,{\frac{0.23+z}{0.64}} ~.
\eeq

The results obtained using this method are shown in
figure~\ref{fig:w=-0.5,D=tanh,SNAP} for SNAP data for the fiducial
model $w=-0.5$. We find that this variable form of $\Delta(z)$ leads
to a slight improvement of results at low redshifts by getting rid of
the small bias which remains in the bootstrap iterative process.  This
improvement of the results is expected especially for the cosmological
models whose equation of state at low redshift is very different as
compared to the $\Lambda$CDM model, which is our initial guess model.

\end{document}